\documentstyle[multicol,eqsecnum,aps,psfig]{revtex}
\renewcommand{\narrowtext}{\begin{multicols}{2}
\global\columnwidth20.5pc\noindent}
\renewcommand{\widetext}{\end{multicols}
\global\columnwidth42.5pc}
\begin{document}
\draft
\preprint{revised 7 May 1999}
\title{Breakdown of a Magnetization Plateau due to Anisotropy
       in Heisenberg Mixed-Spin Chains}
\author{Shoji Yamamoto}
\address
{Department of Physics, Okayama University,
 Tsushima, Okayama 700-8530, Japan}
\author{T$\hat{\mbox o}$ru Sakai}
\address
{Faculty of Science, Himeji Institute of Technology,
 Ako, Hyogo 678-1297, Japan}
\date{18 March 1999}
\maketitle
\begin{abstract}
We discuss the critical behavior of the spin-$(1,\frac{1}{2})$
Heisenberg ferrimagnetic chain in a magnetic field, whose
magnetization curve exhibits a plateau at a third of the full
magnetization.
A bond alternation stabilizes the massive state, whereas an exchange
anisotropy causes the breakdown of the plateau and the onset of a
gapless spin-fluid state, where the transition, lying in the $XY$ but
ferromagnetic region, is of Kosterlitz-Thouless type.
In order to elucidate significant quantum effects, we investigate the
model of classical version as well.
\end{abstract}
\narrowtext

\section{Introduction}\label{S:I}

   Ground-state magnetization curves of quantum spin chains have been
attracting much current interest due to their quantized plateaux as
functions of a magnetic field.
Several years ago Hida \cite{Hida59} revealed that a
spin-$\frac{1}{2}$ ferromagnetic-antiferromagnetic-antiferromagnetic
trimerized chain exhibits a plateau in its magnetization curve at
a third of the full magnetization.
Although it was already familiar that, in the presence of a field,
integer-spin Heisenberg antiferromagnetic chains remain massive from
zero field up to a critical field \cite{Hald64}, yet the
magnetization plateau at a fractional value of the full magnetization
was still met with a surprise.
Since then various low-dimensional quantum spin systems in a field
have been investigated, including polymerized spin chains
\cite{Okam45,Tone17,Tots03,Naka26,Cabr19,Hone}, spin chains with
anisotropy \cite{Saka01} or four-spin exchange coupling \cite{Saka},
and decorated spin ladders \cite{Cabr26,Tand96,Kole}.
Experimental observations \cite{Naru09,Shir48} of quantized
magnetization plateaux have also been reported.
In such circumstances, generalizing the Lieb-Schultz-Mattis theorem
\cite{Lieb07,Affl86}, Oshikawa, Yamanaka, and Affleck (OYA)
\cite{Oshi84} found a criterion for the {\it fractional
quantization}.
They pointed out that quantized plateaux in magnetization curves may
appear under the condition
\begin{equation}
   S_{\rm unit}-m=\mbox{integer}\,,
   \label{E:OYA}
\end{equation}
where $S_{\rm unit}$ is the sum of spins over all sites in the unit
period and $m$ is the magnetization $M$ divided by the number of
the unit cells.

   Mixed-spin chains are the system of all others that stimulates us
in this context.
There exists a large amount of chemical knowledge \cite{Kahn91} on
quantum ferrimagnets.
In an attempt to realize a quasi-one-dimensional ferrimagnetic
system, Gleizes and Verdaguer \cite{Glei73} synthesized a few
bimetallic compounds such as
AMn(S$_2$C$_2$O$_2$)$_2$(H$_2$O)$_3$$\cdot$$4.5$H$_2$O
(A = Cu, Ni, Pd, Pt).
Then numerous chemical explorations \cite{Pei38,Pei47} followed and
various examples of a ferrimagnetic one-dimensional compound were
systematically obtained.
The vigorous experimental research motivated theoretical
investigations into Heisenberg ferrimagnets.
Drillon {\it et al.} \cite{Dril13} pioneeringly carried out
numerical diagonalizations of spin-$(S,\frac{1}{2})$ Heisenberg
Hamiltonians for $S=1$ to $\frac{5}{2}$ and revealed typical
thermodynamic properties of ferrimagnetic mixed-spin chains.
In recent years, quantum ferrimagnets have met with further
theoretical understanding \cite
{Alca67,Pati94,Breh21,Kole36,Nigg31,Yama09,Ono76,Kura62,Yama08,Ivan24,Abol}
owing to various tools such as
field \cite{Alca67,Abol} and
spin-wave \cite{Pati94,Breh21,Yama08,Ivan24} theories,
matrix-product formalism \cite{Kole36,Nigg31}, and
quantum Monte Carlo \cite{Breh21,Yama09,Yama08} and
density-matrix renormalization-group \cite{Pati94,Ono76,Yama08}
techniques.
In particular, their mixed nature, showing both ferromagnetic and
antiferromagnetic aspects \cite{Yama08}, has lately attracted
considerable attention.

 However, little is known about quantum ferrimagnetic behavior in
a magnetic field \cite{Kole36}, especially about magnetization curves
\cite{Kura62}.
Although anisotropy is an interesting and important factor from an
experimental point of view, there exist few arguments on anisotropic
models in a field.
Now, considering the OYA argument and the accumulated chemical
knowledge on ferrimagnetic compounds, the magnetization process of
realistic mixed-spin-chain models arouses our interest all the more
and indeed deserves urgent communication.
In an attempt to serve as guides for further experimental study,
we here consider an alignment of alternating spins $S$ and $s$ in a
field, as described by the Hamiltonian
\begin{equation}
   {\cal H}
    ={\displaystyle\sum_{j=1}^N}
    \left[
     (\mbox{\boldmath$S$}_{j}\cdot\mbox{\boldmath$s$}_{j})_\alpha
    +\delta
     (\mbox{\boldmath$s$}_{j}\cdot\mbox{\boldmath$S$}_{j+1})_\alpha
    -H(S_j^z+s_j^z)
    \right]\,,
   \label{E:H}
\end{equation}
where
$(\mbox{\boldmath$S$}\cdot\mbox{\boldmath$s$})_\alpha
 =S^xs^x+S^ys^y+\alpha S^zs^z$.
We note that even the bond alternation $\delta$ is now experimentally
adjustable \cite{Pei47}.
According to the OYA criterion (\ref{E:OYA}), as $H$ increases from
zero to the saturation field
\begin{equation}
   H_{\rm sat}
    =\frac{1}{2}(1+\delta)
     \left[
      \alpha(S+s)+\sqrt{\alpha^2(S-s)^2+4Ss}
     \right],
   \label{E:Hsat}
\end{equation}
the model (\ref{E:H}) may exhibit quantized plateaux at
$m=\frac{1}{2}$ $(1)$, $\frac{3}{2}$ $(2)$, $\cdots$, $S+s-1$.
Though a multi-plateau problem is a fascinating subject,
we restrict our argument to the simplest case of
$(S,s)=(1,\frac{1}{2})$ in the following.
This is, on the one hand, because we at first aim at understanding
the typical and essential behavior of quantum ferrimagnets in a
field, and, on the other hand, because the low-energy structure of
the model (\ref{E:H}) remains qualitatively the same
\cite{Alca67,Yama08} as long as $S\neq s$.
Then, a plateau is expected at $m=\frac{1}{2}$.
At the Heisenberg point, the ground state of the Hamiltonian
(\ref{E:H}) without field is a multiplet of spin $N/2$ \cite{Lieb49}.
The ferromagnetic excitations, reducing the ground-state
magnetization, exhibit a gapless dispersion relation, whereas the
antiferromagnetic ones, enhancing the ground-state magnetization,
are gapped from the ground state \cite{Yama09}.
Therefore, at the isotropic point, $m$ jumps up to $\frac{1}{2}$ just
as a field is applied and forms a plateau for
$H_{\rm c1}\leq H\leq H_{\rm c2}$ \cite{Kura62}, where
$H_{\rm c1}$ and $H_{\rm c2}$ are the lower and upper critical
fields, equal to $0$ and the antiferromagnetic gap, respectively.
\begin{figure}
\vskip -30mm
\mbox{\psfig{figure=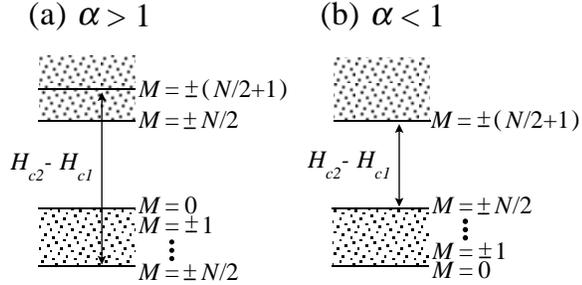,width=90mm,angle=0}}
\vskip 5mm
\caption{Schematic view of the low-energy structure of the
         spin-$(1,\frac{1}{2})$ quantum ferrimagnetic chain with
         anisotropic exchange coupling near the Heisenberg point
         $\alpha=1$:
         (a) the Ising region $\alpha>1$ and
         (b) the $XY$ region $\alpha<1$.}
\label{F:LES}
\end{figure}
\vskip 3mm

   In the presence of exchange anisotropy, the above argument should
be modified, where the $(N+1)$-fold degenerate ground-state multiplet
splits \cite{Alca67,Ono76}, as illustrated in Fig. \ref{F:LES}.
In the Ising region, the ground state is a doublet of $M=\pm N/2$
and therefore $H_{\rm c1}$ remains $0$.
As $\alpha$ increases, $H_{\rm c2}$ comes to be given as
$(1+\delta)\alpha$ and the magnetization curve ends up with a trivial
step.
Thus we take little interest in this region.
In the $XY$ region, on the other hand, the ground state is a singlet
of $M=0$.
Now $H_{\rm c1}$ moves away from $0$ and the plateau shrinks as
$\alpha$ decreases (see Fig. \ref{F:mquant} below).
Here arises a stimulative problem:
how stable the plateau is against the anisotropy and what comes
over the plateau phase?
In this article, we demonstrate that the plateau survives the $XY$
anisotropy in the entire antiferromagnetic region and vanishes in the
ferromagnetic region.
The transition is of Kosterlitz-Thouless (KT) type \cite{Kost81} and
a gapless spin-fluid phase \cite{Hald25} appears instead.
\begin{figure}
\mbox{\psfig{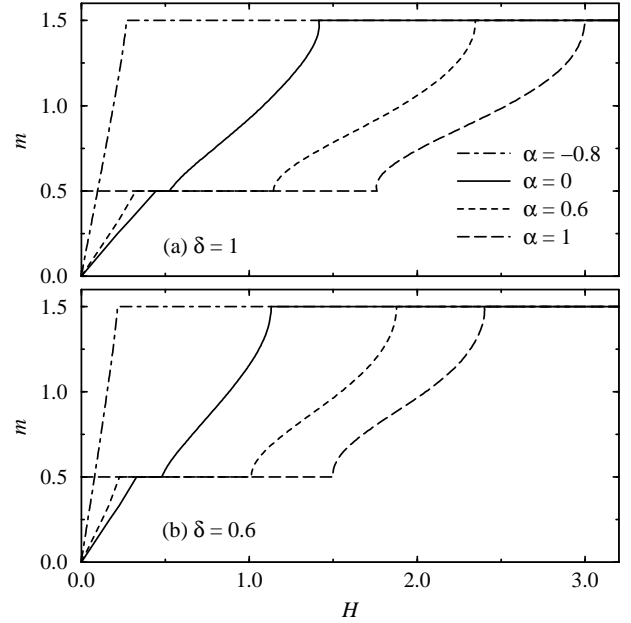}}
\vskip 5mm
\caption{The ground-state magnetization curves for the quantum
         Hamiltonian (1.2) at various values of $\alpha$:
         (a) $\delta=1$ and (b) $\delta=0.6$.}
\label{F:mquant}
\end{figure}
\begin{figure}
\vskip -5mm
\mbox{\psfig{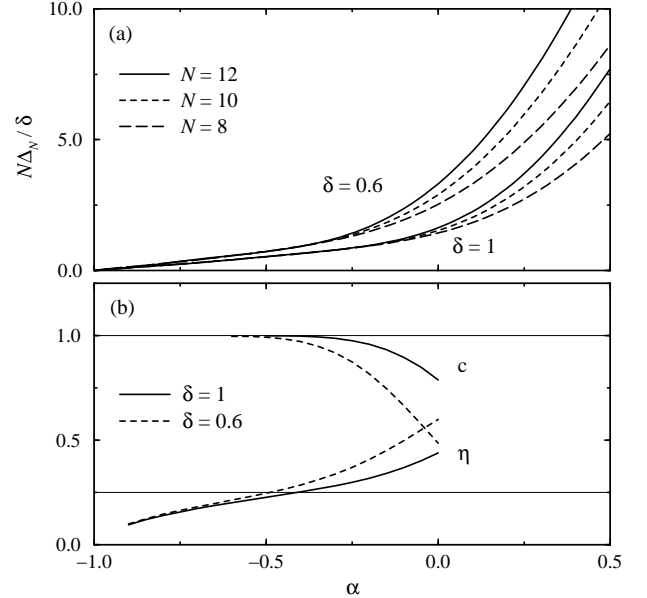}}
\vskip 5mm
\caption{(a) Scaled quantity $N{\mit\Delta}_N$ versus $\alpha$ at
         $\delta=1$ and $\delta=0.6$.
         (b) The central charge $c$ and the critical exponent $\eta$
         versus $\alpha$ in the vicinity of the phase boundary at
         $\delta=1$ and $\delta=0.6$.}
\label{F:CI}
\end{figure}

\section{Scaling Analysis}\label{S:SA}

   We numerically diagonalize finite clusters up to $N=12$ and
analyze the data obtained employing a scaling technique
\cite{Saka01,Saka83}.
Suppose a field is applied to the cluster of $N$ unit cells, a
magnetization, let us say, $M$, is induced in the ground state.
In this sense we represent a field as a function of $N$ and $M$:
$H(N,M)$.
Even though $M$, as well as $N$, is given, $H(N,M)$ is not in general
unique.
The upper and lower bounds of $H(N,M)$ are, respectively, given by
\begin{eqnarray}
   &&
   H_+(N,M)=E(N,M+1)-E(N,M),
   \label{E:H+}
   \\
   &&
   H_-(N,M)=E(N,M)-E(N,M-1),
   \label{E:H-}
\end{eqnarray}
where $E(N,M)$ is the lowest energy in the subspace labeled $M$ of
the Hamiltonian (\ref{E:H}) without the Zeeman term.
If the system is massive at the sector labeled $M$, $H_\pm(N,M)$
should approach different values $H_\pm(m)$, respectively, as
$N\rightarrow\infty$, which can be estimated by the Shanks'
extrapolation \cite{Shan01}.
In the critical system, on the other hand, $H_\pm(N,M)$ should
converge to the same value as \cite{Card85,Affl46}
\begin{equation}
   H_{\pm}(N,M)
    \sim H(m) \pm \frac{\pi v_{\rm s}\eta}{N}
    \ \ (N\rightarrow\infty)\,,
    \label{E:CFT}
\end{equation}
where $v_{\rm s}$ is the sound velocity and $\eta$ is the critical
index defined as
$\langle\sigma^+_0\sigma^-_r\rangle \sim (-1)^r r^{-\eta}$
with a relevant spin operator $\sigma$, which may here be a certain
linear combination of $\mbox{\boldmath$S$}$ and
$\mbox{\boldmath$s$}$.

   In Fig. \ref{F:mquant} we show thus-obtained thermodynamic-limit
magnetization-versus-field curves, where we smoothly interpolate the
raw data $H(m)$ for the sake of guiding eyes.
We might expect that the bond alternation simply makes the plateau
grow because the magnetization curve becomes stepwise as
$\delta\rightarrow 0$.
However, this naive idea is not true in general.
In the vicinity of the Ising limit $\alpha\rightarrow\infty$, the
plateau length behaves as $(1+\delta)\alpha$ and thus the bond
alternation makes the plateau shrink.
Around the Heisenberg point $\alpha=1$, this picture seems to be
still valid in part but the precise scenario is not so simple.
At the Heisenberg point, for example, the antiferromagnetic
excitation gap, that is, the gap between the ground state and the
lowest level in the subspace with $M=N/2+1$, is not a monotonic
function of $\delta$ (Table \ref{T:AFgap}).
On the other hand, near the $XY$ point $\alpha=0$, the plateau seems
to grow monotonically with the bond alternation.

   Once $\delta$ is given, the plateau length is monotonically
reduced with the decrease of $\alpha$.
The system is gapless at every sector of the Hilbert space in the
ferromagnetically ordered region $\alpha\leq -1$ and is thus supposed
to encounter a phase transition going through the $XY$ region
$-1<\alpha<1$.
It is surprising that the plateau still exists at the $XY$ point.
We will show later that such a stable plateau is peculiar to quantum
spins, while, for classical spins, only a slight anisotropy of $XY$
type breaks the plateau.
\begin{figure}
\mbox{\psfig{figure=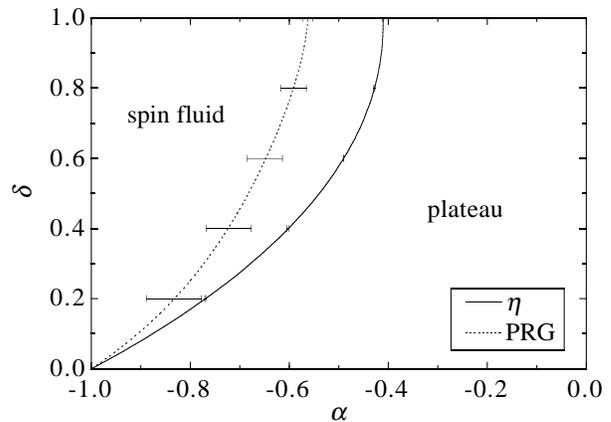,width=80mm,angle=0}}
\vskip -3mm
\caption{Phase diagram of the spin-$(1,\frac{1}{2})$ quantum
         ferrimagnetic chain (1.2) at the absolute zero temperature.
         The phase boundary determined by the critical index $\eta$
         is shown by a solid line, whereas the PRG estimate by a
         dotted line.
         The dominant error for the PRG result occurs in
         extrapolating $\alpha_{\rm c}(N,N+2)$ to the
         $N\rightarrow\infty$ limit rather than originates from the
         numerical diagonalization.}
\label{F:PhD}
\end{figure}
\vskip 3mm

   The plateau length ${\mit\Delta}_N=H_+(N,M)-H_-(N,M)$ is a
relevant order parameter to detect the phase boundary. 
The scaling relation (\ref{E:CFT}) suggests that ${\mit\Delta}_N$
should be proportional to $1/N$ in the critical system.
We plot in Fig. \ref{F:CI}(a) the scaled quantity $N{\mit\Delta}_N$
as a function of $\alpha$.
$N{\mit\Delta}_N$ looks independent of $N$ beyond a certain value of
$\alpha$, showing an aspect of the KT transition.
The central charge $c$ of the critical phase can be extracted from
the scaling relation of the ground-state energy:
\begin{equation}
   \frac{E(N,M)}{N}
    \sim \varepsilon(m)-\frac{\pi cv_{\rm s}}{N^2}
    \ \ (N\rightarrow\infty) \,.
\label{E:GS}
\end{equation}
Due to the small correlation length \cite{Pati94,Breh21} of the
present system, we can directly and precisely estimate $v_{\rm s}$
from the dispersion curves.
In Fig. \ref{F:CI}(b) we plot $c$ versus $\alpha$ and find that $c$
approaches unity as the system goes toward the critical region.
Assuming the asymptotic formula
${\mit\Delta}_N\sim 2\pi v_{\rm s}\eta/N$, we can further evaluate
the critical exponent $\eta$, which is also shown in Fig.
\ref{F:CI}(b).
Figure \ref{F:CI} fully convinces us of the KT universality of this
phase transition.
The phase boundary is obtained by tracing the points of
$\eta=\frac{1}{4}$ \cite{Schu72} and is shown in Fig. \ref{F:PhD} by
a solid line.
On the other hand, we have another numerical tool, the
phenomenological renormalization-group (PRG) technique \cite{Nigh61},
to determine the phase boundary.
At each $\delta$, the PRG equation
\begin{equation}
   (N+2){\mit\Delta}_{N+2}(\alpha,\delta)
    =N{\mit\Delta}_N(\alpha,\delta)\,,
\label{E:PRG}
\end{equation}
gives size-dependent fixed points $\alpha_{\rm c}(N,N+2)$.
$\alpha_{\rm c}(N,N+2)$ is well fitted to a linear function of
$1/(N+1)$ in the vicinity of $\delta=1$, whereas, as
$\delta\rightarrow 0$, the linearity becomes worse and thus the
uncertainty in the $N\rightarrow\infty$ extrapolation increases.
Just for reference, the thus-obtained phase boundary is also shown in
Fig. \ref{F:PhD} by a dotted line, which is somewhat discrepant from
the highly accurate estimate based on $\eta$.
The PRG equation applied to gapful-to-gapful phase transitions yields
an accurate solution, to be sure, but, for transitions to a gapless
phase, including those of KT type, the PRG analysis is likely to miss
the correct solution due to essential corrections to the scaling
law (\ref{E:CFT}), overestimating the gapful-phase region
\cite{Nomu23,Okam79}.
The present PRG solution may still be recognized as the lower
boundary of $\alpha_{\rm c}$.
\begin{figure}
\vskip 3mm
\mbox{\psfig{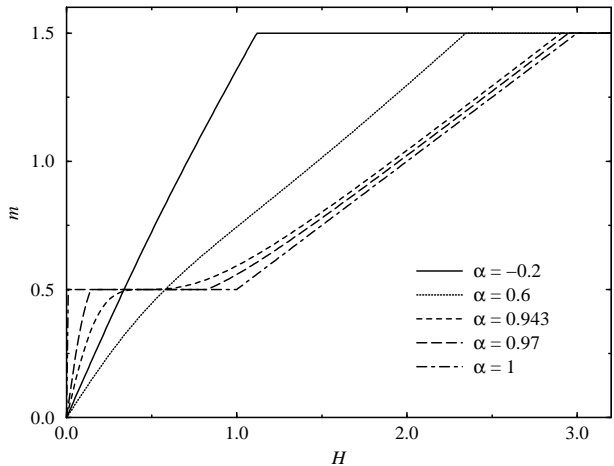}}
\vskip 5mm
\caption{The ground-state magnetization curves for the classical
         Hamiltonian (1.2) with $\delta=1$ at various values of
         $\alpha$.}
\label{F:mclass}
\end{figure}

\section{Sublattice Magnetizations}\label{S:R}

\begin{figure}
\vskip 2mm
\mbox{\psfig{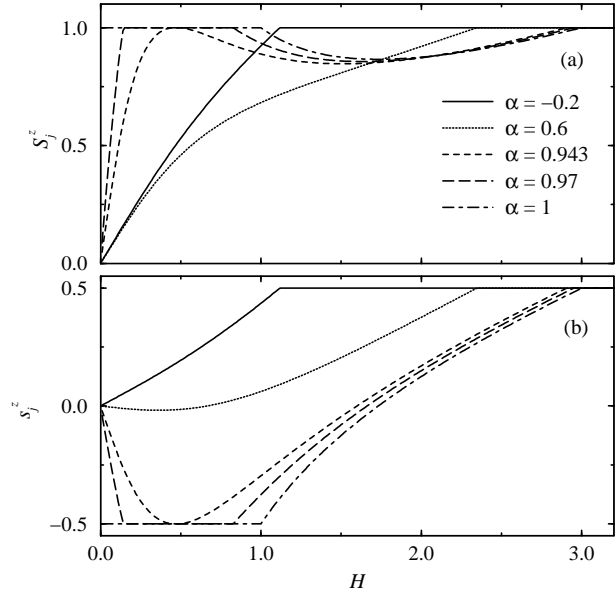}}
\vskip 5mm
\caption{The ground-state sublattice magnetizations per unit cell as
         functions of a field for the classical Hamiltonian (1.2)
         with $\delta=1$ at various values of $\alpha$:
         (a) the larger spin $S=1$ and
         (b) the smaller spin $\frac{1}{2}$.}
\label{F:submclass}
\end{figure}

\begin{figure}\mbox{\psfig{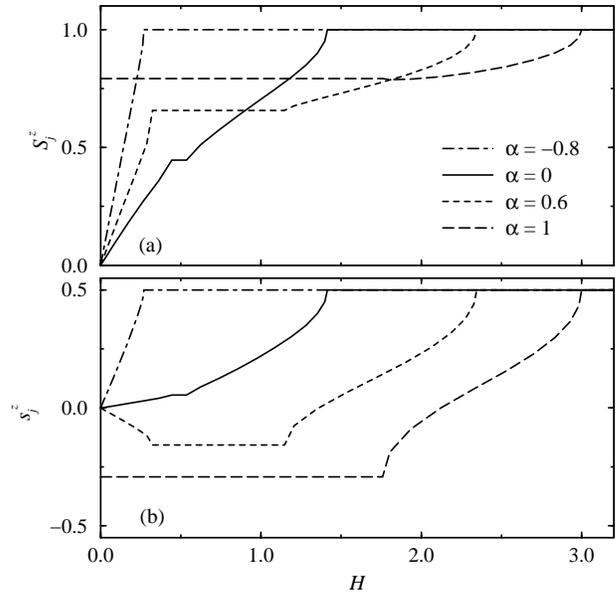}}
\vskip 5mm
\caption{The ground-state sublattice magnetizations per unit cell as
         functions of a field for the quantum Hamiltonian (1.2) with
         $\delta=1$ at various values of $\alpha$:
         (a) the larger spin $S=1$ and
         (b) the smaller spin $\frac{1}{2}$.}
\label{F:submquant}
\end{figure}

   In an attempt to elucidate how much effect quantum fluctuations
have on the stability of the plateau, we investigate the Hamiltonian
(\ref{E:H}) of classical version as well, where
$\mbox{\boldmath$S$}_j$ and $\mbox{\boldmath$s$}_j$ are classical
vectors of magnitude $1$ and $\frac{1}{2}$, respectively.
We show in Fig. \ref{F:mclass} the classical magnetization curves.
We note that the classical model also exhibits a plateau at
$m=\frac{1}{2}$.
The magnetization curves in the Ising region are not so far from the
quantum behavior, though we have not shown them explicitly.
However, the classical plateau can hardly stand the anisotropy of
$XY$ type.
In this context, it is interesting to observe sublattice
magnetizations separately.
We show in Fig. \ref{F:submclass} the configuration of each classical
spin as a function of a field.
The classical plateau is nothing but a N\'eel-ordered state.
In other words, without the fully ordered staggered magnetization,
classical spins could not form a magnetization plateau.
On the other hand, Fig. \ref{F:submquant} shows that quantum spins
can form a magnetization plateau with any combination of sublattice
magnetizations.
It is the case with the quantum model as well that sublattice
magnetizations themselves freeze while going through the plateau.
However, as long as the $XY$ exchange interaction exists, they are in
general reduced from the full values $1$ and $-\frac{1}{2}$,
respectively.
It is quantum fluctuations that stabilize the plateau with
unsaturated sublattice magnetizations.
\vskip -1mm
\vskip 2mm

   One more interesting observation on the quantum spin configuration
is that the collapse of the staggered order in $z$ direction neither
coincides with the $XY$ point nor results in the disappearance of the
plateau.
The $z$-direction spin correlations between the two sublattices
turn ferromagnetic before the model reaches the $XY$ point.
Here let us be reminded of the mixed nature \cite{Yama08} of quantum
ferrimagnets.
Because of the coexistent elementary excitations of different types,
the specific heat exhibits a Schottky-like peak in spite of the
initial ferromagnetic behavior at low temperatures, whereas the
susceptibility-temperature product shows both increasing and
decreasing behaviors as functions of temperature.
The present phenomenon, a massive state in the ferromagnetic
background, might also be recognized as a combination of
ferromagnetic and antiferromagnetic features.

\section{Summary and Discussion}\label{S:SD}

   We have investigated the critical behavior of anisotropic
Heisenberg mixed-spin chains in a field.
The model shows an anisotropy-induced transition of KT type between
the plateau and spin-fluid phases, whose phase boundary lies in the
ferromagnetic-coupling region.
Though we have restricted our argument to the case of
$(S,s)=(1,\frac{1}{2})$, qualitatively the same scenario may be
expected in higher-spin cases, where multi-plateau phases are
possible with the assistance of bond alternation \cite{Yama}.

   While our scaling analysis is highly accurate, it is subtle
whether or not the plateau still exists at the $XY$ point.
Therefore, any other argument would be helpful in understanding
further the numerical findings obtained.
Let us consider a spin-$\frac{1}{2}$
ferromagnetic-antiferromagnetic-antiferromagnetic trimerized chain
\begin{equation}
 {\cal H}
  ={\displaystyle\sum_{j=1}^N}
  \left[
  -\gamma
   (\mbox{\boldmath$\sigma$}^a_{j  }
    \cdot
    \mbox{\boldmath$\sigma$}^b_{j  })_\alpha
  +(\mbox{\boldmath$\sigma$}^b_{j  }
    \cdot
    \mbox{\boldmath$\sigma$}^c_{j  })_\alpha
  +(\mbox{\boldmath$\sigma$}^c_{j  }
    \cdot
    \mbox{\boldmath$\sigma$}^a_{j+1})_\alpha
  \right]\,,
  \label{E:HFAA}
\end{equation}
which can be regarded as the Heisenberg ferrimagnet of our interest
in the $\gamma\rightarrow\infty$ limit.
Such a replica-model approach is quite useful \cite{Hida07} in
studying low-dimensional quantum magnetism.
Introducing the Jordan-Wigner spinless fermions via
\begin{equation}
   \lambda_j^\dagger
    ={\sigma^\lambda_j}^+
     {\rm exp}
     \biggl[
      -{\rm i}\pi\sum_{l=1}^{j-1}
       {\sigma^\lambda_l}^+
       {\sigma^\lambda_l}^-
     \biggr]
   \ \ (\lambda=a,b,c)\,,
\end{equation}
we replace the Hamiltonian (\ref{E:HFAA}) by
\begin{equation}
 {\cal H}
  =\sum_{j=1}^N
   \left[
    (a_j,b_j)_{-\gamma,\alpha}
   +(b_j,c_j)_{1,\alpha}
   +(c_j,a_{j+1})_{1,\alpha}
   \right]\,,
  \label{E:HFAASF}
\end{equation}
where
$4(a,b)_{\gamma,\alpha}
 =2\gamma(a^\dagger b+b^\dagger a)
 +\alpha(2a^\dagger a-1)(2b^\dagger b-1)$.
\begin{figure}
\vskip -5mm
\mbox{\psfig{figure=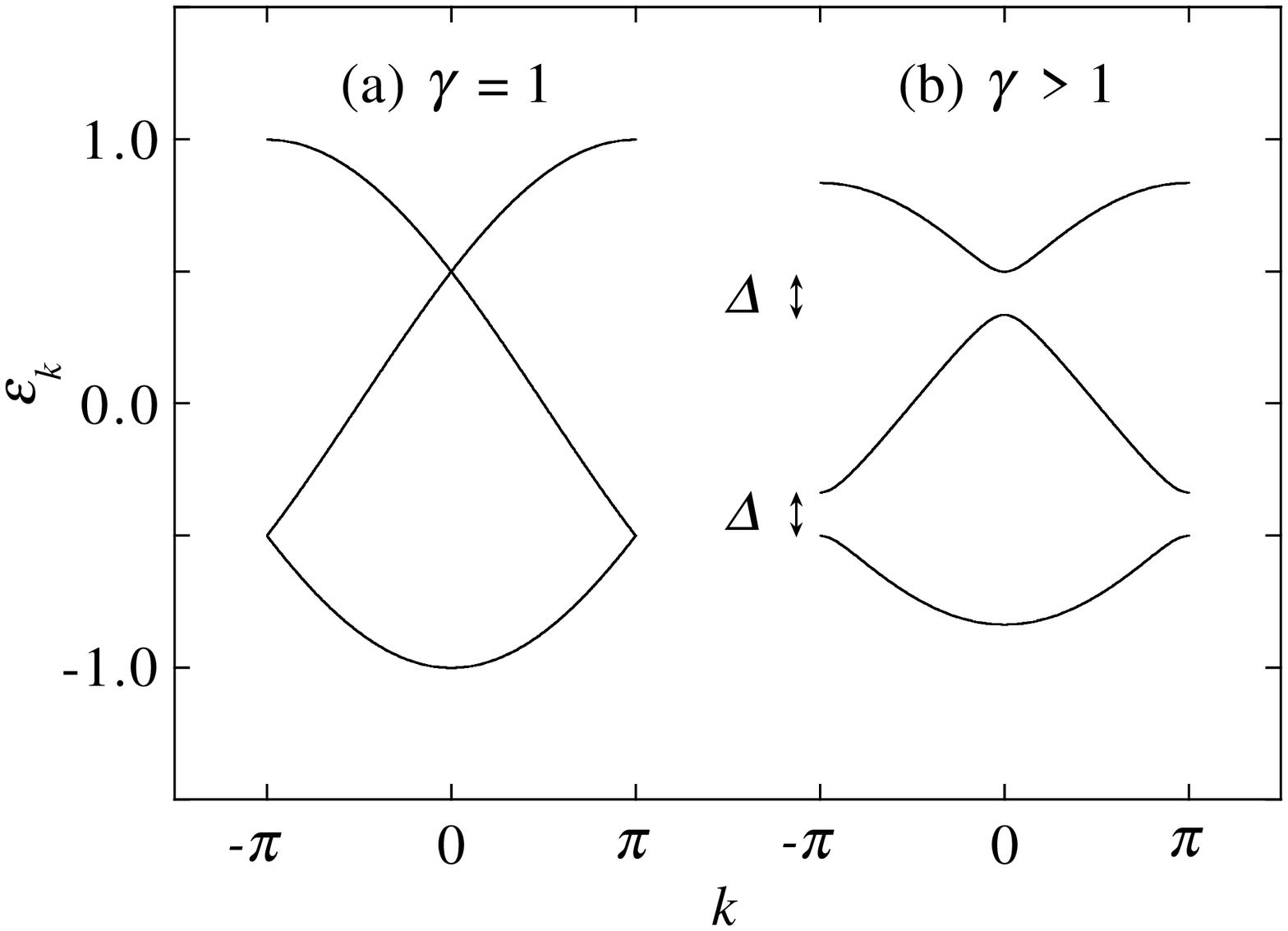,width=80mm,angle=0}}
\vskip -3mm
\caption{Dispersion relations of the spin-$\frac{1}{2}$ trimerized
         chain (4.1) at the $XY$ point $\alpha=0$.
         (a) $\gamma=1$.
         There is no gap in the excitation spectrum.
         (b) $\gamma>1$.
         There open up gaps at the sectors of $\frac{1}{3}$ and
         $\frac{2}{3}$ band filling, where
         $2\Delta=3\gamma-(\gamma^2+8)^{1/2}$.}
\label{F:dspFAA}
\end{figure}
\begin{figure}
\mbox{\psfig{figure=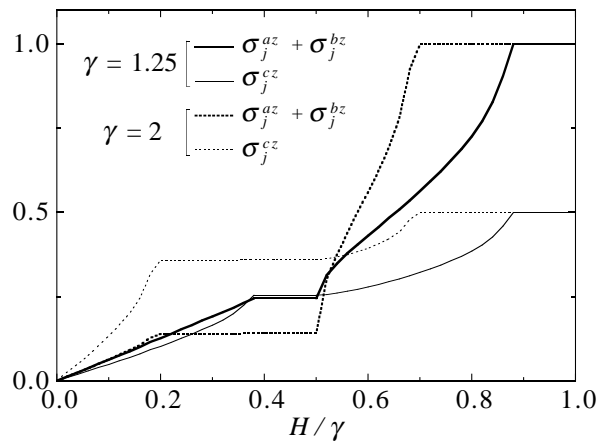,width=80mm,angle=0}}
\vskip -3mm
\caption{The ground-state sublattice magnetizations per unit cell as
         functions of a field of the trimerized spin-$\frac{1}{2}$
         chain (4.1) at the $XY$ point $\alpha=0$ for $\gamma=1.25$
         and $\gamma=2$.}
\label{F:submFAA}
\end{figure}
\vskip 3mm

   Now we focus our interest on the $XY$ point $\alpha=0$.
After the Fourier transformation, we obtain the equation to
determine the single-particle excitation spectrum as
\begin{equation}
   \varepsilon_k^3-(\gamma^2+2)\varepsilon_k-2\gamma\cos k=0\,.
\end{equation}
The resultant dispersion relation is qualitatively different
according as $\gamma=1$ or not, as illustrated in Fig.
\ref{F:dspFAA}.
At $\gamma=1$, which is not large enough to let ferromagnetically
coupled neighboring spins construct spin $1$'s, there is no gap in
the excitation spectrum.
However, as $\gamma$ increases, gaps open up at the sectors of
$\frac{1}{3}$ and $\frac{2}{3}$ band filling and this scenario
remains qualitatively unchanged in the whole region $\gamma>1$.
Noting the relation between the magnetization and the band filling,
\begin{equation}
   M=N_{\rm occ}-\frac{3N}{2}\,,
\end{equation}
where $N_{\rm occ}$ is the number of occupied states, we are allowed
to expect magnetization plateaux at $m=\pm\frac{1}{2}$.
The inclusion of the bond alternation $\delta$ results in the
enhancement of the gap, which is consistent with Fig. \ref{F:mquant}.
Qualitatively the same scenario is available for a
ferromagnetic-ferromagnetic-antiferromagnetic trimerized chain, as
was pointed out by two pioneering authors \cite{Zasp13,Okam39}.
The present analysis is not strictly comparable to the original
argument unless $\alpha=1$.
However, the nonvanishing gap in the $\gamma\rightarrow\infty$ limit
may be a qualitative evidence for the existence of the plateau at the
$XY$ point in the original model (\ref{E:H}).
We further show in Fig. \ref{F:submFAA} the sublattice magnetizations
in the ground state of the replica model with $\alpha=0$ as functions
of a field at a few values of $\gamma>1$.
We are convinced all the more that the N\'eel order has already
disappeared and both the spins $1$ and $\frac{1}{2}$ have the
same-sign $z$ components at the $XY$ point.

   In recent years, a massive-to-spin-fluid phase transition of KT
type has been given a great deal of attention
\cite{Bote14,Soly80,Alca96,Scho93,Yaji34,Malv85,Neir03} in the
context of Haldane's conjecture \cite{Hald64}.
In such cases the critical point never goes beyond the $XY$ point.
The magnetization plateau in our argument should be distinguished
from the gap immediately above the ground state, to be sure, but,
compared with Haldane's scenario \cite{Schu72}, the present
observation looks novel and is fascinating to be further studied.
There may be a new mass-generation mechanism peculiar to quantum
mixed-spin chains, other than the valence-bond picture \cite{Affl99}.
Quite recently Okamoto and Kitazawa \cite{Okam} have reported that
the magnetization plateau of the spin-$\frac{1}{2}$ trimerized chain
which is closely related with the present model also disappears in
the $XY$ ferromagnetic region.
We hope that our investigation, combined with such an argument from
a different viewpoint, will contribute toward revealing the
possibly novel scenario for the breakdown of quantized plateaux.

\acknowledgments

   It is a pleasure to thank H.-J. Mikeska and U. Schollw\"ock for
helpful discussions.
This work was supported by the Japanese Ministry of Education,
Science, and Culture through Grant-in-Aid No. 09740286 and by the
Okayama Foundation for Science and Technology.
The numerical computation was done in part using the facility of the
Supercomputer Center, Institute for Solid State Physics, University of
Tokyo.

\begin{table}
\vskip 5mm
\caption{The antiferromagnetic excitation gap ${\mit\Delta}$ as a
         function of $\delta$ at the Heisenberg point $\alpha=1$.}
\begin{tabular}{ccccccc}
   $\delta$    &   $1$    &  $0.8$   &  $0.6$   &  $0.4$   &
                  $0.2$   &   $0$         \\
${\mit\Delta}$ & $1.7591$ & $1.6042$ & $1.4986$ & $1.4500$ &
                 $1.4558$ & $\frac{3}{2}$ \\
\end{tabular}
\label{T:AFgap}
\end{table}
\widetext
\end{document}